\documentclass[12pt]{article}
\usepackage[T1]{fontenc}
\usepackage{a4wide}
\usepackage{geometry}
\usepackage{amsmath}
\usepackage[dvips]{graphics}

\makeatletter

\newcommand{\LyX}{L\kern-.1667em\lower.25em\hbox{Y}\kern-.125emX\spacefactor1000}

\makeatother

\begin{document}

\noindent March, 1998\hfill{}DFUB-98-04 

\vspace{0.8cm}

{\centering \textbf{\large TRUNCATED CONFORMAL SPACE AT \( c=1 \), NONLINEAR INTEGRAL
EQUATION AND QUANTIZATION RULES FOR MULTI-SOLITON STATES}\\
\textbf{\large \vspace{0.8cm}}\large \par}

{\centering {\large G. Feverati, F. Ravanini and G. Takács}\\
\vspace{0.4cm}\par}

{\centering \emph{Sez. INFN and Dip. di Fisica - Univ. di Bologna}\\
\emph{Via Irnerio 46, I-40126 BOLOGNA, Italy}\\
\emph{\vspace{1cm}}\par}

{\centering \textbf{Abstract}\par}

{\centering \vspace{0.2cm}\par}

{\footnotesize We develop Truncated Conformal Space (TCS) technique for perturbations
of \( c=1 \) Conformal Field Theories. We use it to give the first numerical evidence
of the validity of the non-linear integral equation (NLIE) derived from light-cone
lattice regularization at intermediate scales. A controversy on the quantization
of Bethe states is solved by this numerical comparison and by using the locality
principle at the ultraviolet fixed point. It turns out that the correct quantization
for pure hole states is the one with half-integer quantum numbers originally
proposed by Fioravanti et al. \cite{noi}. Once the correct rule is imposed, the agreement
between TCS and NLIE for pure hole states turns out to be impressive.}{\footnotesize \par}

\section{Introduction}

The scaling functions of 1+1 dimensional integrable models on the cylinder have
proven to be a very useful non-perturbative tool of investigation of their finite
size effects, in particular of the renormalization flow properties of the vacuum
as well as of the excited states. Various methods have been proposed for the
calculation of these important quantities, as the Truncated Conformal Space
(TCS) method \cite{yurzam} or the Thermodynamic Bethe Ansatz (TBA) \cite{YY,Zam-tba1}. A very promising method
is the one introduced some years ago by Destri and de Vega \cite{ddv-92,ddv-95} (similar methods
were independently introduced in Condensed Matter Physics by other authors \cite{klumper}).
It consists in defining a lattice model and give evidence that its continuum
limit reproduces the sine-Gordon QFT (sG) \cite{ddv-87}, then deriving a non-linear integral
equation (NLIE) which is basically the continuum limit of the Bethe equations.
There is one such equation for each excitation in the spectrum of the physical
states. The problem is to give a general rule to write down the NLIE corresponding
to a state characterized by a certain distribution of Bethe roots and holes.
This problem, namely to find the NLIE for excited states, was first addressed,
in a QFT context, in \cite{noi} which discussed the case of pure hole excitations. The
general setup of excited states was described in detail later in \cite{ddv-97}. The solution
of the NLIE, the so-called \emph{counting function}, is a central object in
Bethe Ansatz approach to integrable QFT from which it is possible to reconstruct
the eigenvalues of all the local conserved currents of the theory put on a cylinder
of circumference \( L \).

The aim of this paper is to present a numerical comparison between the NLIE
and TCS. We develop a TCS for \( c=1 \) CFT perturbed by its \( \cos \beta \phi  \) operator, that defines
the sG model as perturbation of a CFT. We compare TCS data against scaling functions
computed from numerical integration of NLIE and find a very good agreement,
especially in the attractive regime, but also in the repulsive one. This agreement,
however, only shows up for a specific choice of the quantization rule for the
Bethe Ansatz states which is different from the one reported in \cite{ddv-97} and instead
agrees with the one made in ref.\cite{noi}. This leads to the observation that the NLIE
produces in fact more states than those present in the sG Hilbert space. It
is known that the relation among the Hilbert space of \( c=1 \) CFT, the sG model and
the massive Thirring (mTh) model is a quite delicate issue \cite{kl-me}. It is out of the
scope of the present letter to investigate this very important question. We
intend to return to it in a more extensive paper where a more detailed analysis
of the results presented here will be given \cite{noi3}. It is important to keep in mind
that the NLIE (and in general Bethe Ansatz methods) are only able, at present,
to reproduce states with even topological charge (i.e. even number of solitons
minus antisolitons) in sG model.

\section{The NLIE and its properties}

The Lagrangian of sG theory is

\[
{\cal L}=\int dx\left( \frac{1}{2}\partial _{\mu }\Phi \partial ^{\mu }\Phi +\frac{{\cal M}^{2}}{\beta ^{2}}:\cos \left( \beta \Phi \right) :\right) .\]
Often it is convenient to use the parameter \( p=\frac{\beta ^{2}}{8\pi -\beta ^{2}} \) or the 6-vertex anisotropy \( \gamma =\frac{\pi }{p+1} \). \footnote{
In terms of \( p \), \( p=1 \) is the free fermion point, and \( p=\frac{1}{k}\; ,\; k=1,2,\ldots  \) are the thresholds where a
new breather appears. \( p<1 \) corresponds to the attractive and \( p>1 \) to the repulsive
regime.
} 

We do not reproduce here the deduction of the NLIE from the lattice Bethe equations.
Although there are subtle remarks to do on all the derivations presented so
far \cite{ddv-95,ddv-97}, this aspect is out of the scope of this short letter and will be dealt
with in a forthcoming longer publication \cite{noi3}. Here we are content to present the
final equation and explain how it can be used to investigate the finite size
corrections to energy-momentum of the excited states.

The NLIE for the sG/mTh model reads

\begin{equation}
\label{nlie}
Z(\lambda )=l\sinh \lambda +g(\lambda |\lambda _{j})+2\Im \mathrm{m}\int ^{+\infty }_{-\infty }dxG(\lambda -x-i\eta )\log (1+(-1)^{\delta }e^{iZ(x+i\eta )})
\end{equation}
where \( \eta  \) is a positive free parameter, small enough not to reach any singularity
of the \textit{counting function} \( Z \) off the real axis. With such prescription,
eq. (\ref{nlie}) is totally independent of the choice of \( \eta  \) by Cauchy's theorem. \( l \) is
related to the circumference of the space-time cylinder \( L \) (the spatial volume)
by \( l=ML \), where \( M \) is the mass scale of the theory, chosen to be equal to the soliton
mass. The kernel \( G(\lambda ) \) is given by
\[
G(\lambda )=\int ^{+\infty }_{-\infty }dk\, e^{ik\lambda }\frac{\sinh \frac{(p-1)k\pi }{2}}{\sinh \frac{pk\pi }{2}\, \cosh \frac{k\pi }{2}}.\]
The source term \( g(\lambda |\lambda _{j}) \) is specific to the state considered and vanishes in the case
of the vacuum, which in Bethe Ansatz language corresponds to a sea of real roots. 

The counting function \( Z(\lambda ) \) has the property that any Bethe root satisfies 
\begin{equation}
\label{quantrel}
e^{iZ(\lambda _{j})}=-(-1)^{\delta }.
\end{equation}
The converse is not true, i.e. there are values \( \lambda _{h} \) for which \( e^{iZ(\lambda _{h})}=-(-1)^{\delta } \) that are \emph{not}
solutions of the Bethe equations. Such locations on the real axis are called
\emph{holes.} 

Excited states are seen, in a typical Bethe Ansatz spirit, as different root
configurations for a system of Bethe Ansatz equations and are given by creating
some holes in the vacuum sea and adding some complex roots. Holes and complex
roots are \( O(1) \) in number, while the real roots of the vacuum sea are \( O(N) \) (\( N \) being
the number of sites), and tend to infinity in the continuum limit. The number
\( \delta  \) is equal to the number of roots modulo 2 on the lattice. 

For a state having \( H \) holes at positions \( \lambda _{h} \) on the real axis, \( N_{c} \) complex roots
\( \lambda _{c} \) (of \emph{close} type) such that \( |\Im \mathrm{m}\, \lambda _{c}|<\pi \min (1,p) \), \( N_{w} \) complex roots (of \emph{wide} type)
with \emph{\( |\Im \mathrm{m}\, \lambda _{c}|\geq \pi \min (1,p) \)} and \( N_{s} \) special roots/holes \( \lambda _{s} \):
\[
g(\lambda |\lambda _{j})=\sum _{h=1}^{H}\chi (\lambda -\lambda _{h})-\sum _{c=1}^{N_{c}}\chi (\lambda -\lambda _{c})-\sum _{w=1}^{N_{w}}\chi _{II}(\lambda -\lambda _{w})-2\sum ^{N_{s}}_{s=1}\chi (\lambda -\lambda _{s})\]
where \( \lambda _{j} \) denotes the set of all parameters (\( j=h,c,w,s \)). The function \( \chi (\lambda ) \) is the odd primitive
of \( G(\lambda ) \) and
\[
\chi _{II}(\lambda )=\chi (\lambda )+\mathrm{sign}(p-1)\, \chi (\lambda -i\pi \, \min (1,p)\, \mathrm{sign}(\Im m\, \lambda ))\, .\]
is the so-called \emph{second determination} of the function \( \chi  \). The interested
reader can find more details in \cite{ddv-97}.

The parameters \( \lambda _{j} \) are determined by the Bethe conditions
\begin{equation}
\label{quant}
Z(\lambda _{j})=2\pi I_{j}\qquad ,\qquad I_{j}\in \mathbf{Z}+\frac{1+\delta }{2}
\end{equation}
which are the continuum analogs of the logarithms of the Bethe equations. Any
root of the Bethe equations, as well as any hole, satisfies (\ref{quant}) and its position
is therefore fixed by giving its \emph{Bethe quantum number} \( I_{j} \) which is \textit{half-integer}
for \( \delta =0 \) and \textit{integer} for \( \delta =1 \).

The number \( N_{s} \) of special roots/holes \( \lambda _{s} \), for which \( Z'(\lambda _{s})<0 \), as introduced in \cite{ddv-97}, need
\emph{not} to be specified a priori for a state. Indeed one can easily convince
oneself that in the deep infrared limit \( l\rightarrow \infty  \), \( Z'(\lambda ) \) is always positive and these objects
never appear. Their appearance in some states for values of \( l \) smaller than a
certain \( l_{0} \) is dictated by the fact that sometimes the logarithm in the convolution
term of the equation fails to remain in its fundamental domain, and an analytic
continuation imposes to modify the equation itself by adding new terms to the
source. The number \( H_{eff}=H-2N_{s} \) is invariant for a given state as the spatial volume \( l \)
varies and coincides with the number of dressed \emph{solitonic particles} in
the state.

The number of the various species of roots and holes in a configuration is not
arbitrary but rather is subject to constraints originating from an analysis
of the asymptotics of the lattice counting function. These constraints can be
recast in a form that does not contain the number of real roots, and therefore
their validity can be extended to the continuum limit, in a form that we call
the \emph{counting equation} \cite{ddv-97}
\[
H_{eff}=2S+M_{c}+2M_{w}\, \Theta (p-1)\]
where \( \Theta (x) \) denotes the usual step function and \( S \) is the total spin of the Bethe
Ansatz system, which can be identified with half of the sG topological charge
\( Q \).

Once the function \( Z(\lambda ) \) is evaluated as a solution of the NLIE, it can be used to
compute the energy-momentum \( (E,P) \) (or in principle the eigenvalue of any other local
integral of motion) 
\[
P^{\pm }=E\pm P=z(\lambda |\lambda _{j})-2M\Im \mathrm{m}\int _{-\infty }^{+\infty }\frac{d\lambda }{2\pi }e^{\pm \lambda }\log (1+(-1)^{\delta }e^{iZ(\lambda +i\eta )})\]
where
\[
z(\lambda |\lambda _{j})=M\left( \sum ^{N_{h}}_{h=1}e^{\pm \lambda _{h}}-\sum _{c=1}^{N_{c}}e^{\pm \lambda _{c}}+\Theta (1-p)\sum _{w}e^{\pm \lambda _{w}}\left( 1+e^{\mp i\pi (p+1)\mathrm{sign}(\Im m\lambda _{w})}\right) -2\sum ^{N_{s}}_{s=1}e^{\pm \lambda _{s}}\right) \]

In particular, \( P^{\pm }(L) \) for \( L\rightarrow \infty  \) reproduces the asymptotic particle content of the state.
For \( L\rightarrow 0 \) instead, it is expected to be related to the central charge \( c \) and conformal
dimensions \( \Delta ^{\pm } \) of the ultraviolet CFT state by \( P^{\pm }=-\frac{\pi }{6L}(c-24\Delta ^{\pm }) \).

Notice that eq. (\ref{nlie}) does not coincide with the one reported in \cite{ddv-97}. The argument
of one of logarithmic terms in \cite{ddv-97} differs by a phase \( (-1)^{\delta } \). Our choice gives the
correct infrared asymptotic \( O(e^{-l}) \) for the IR asymptotics of finite size effects,
as it must be in a massive theory. The expression in \cite{ddv-97} would instead give a
rapidly oscillating integrand, which gives a contribution that can be (after
tedious calculations) proven to be only \( O(1/l) \) in the IR limit. A careful analysis
of logarithmic branches confirms that our choice is the correct one. We shall
report the details about this point in a later publication \cite{noi3}.

\section{\label{pureholes_section}UV behaviour of multi-soliton states}

We consider sG theory as the perturbation of a \( c=1 \) massless free boson \( \Phi  \) compactified
on circle of radius \( R \) by a potential \( V=\frac{g}{2}:(e^{i\beta \Phi }+e^{-i\beta \Phi }): \), where \( g=\frac{{\cal M}^{2}}{\beta ^{2}} \).The interaction term has conformal
dimensions \( \Delta _{V}=\Delta _{V}^{\pm }=\frac{\beta ^{2}}{8\pi }=\frac{p}{p+1} \) and becomes marginal when \( \beta ^{2}=8\pi  \) which corresponds to \( p=\infty  \). With the convention\footnote{
Note that this convention is different from the one usually adopted in the literature
(see \cite{kl-me} and references therein), where the compactification radius is \( 1/R \) and the
elementary sine-Gordon field is the dual of the conformal scalar at the UV fixed
point. In our convention, the conformal scalar field is the UV limit of the
elementary sine-Gordon field.
} adopted here, the primary states of the unperturbed CFT, have left and right
conformal dimensions
\[
\Delta ^{\pm }_{n,m}=\frac{1}{2}\left( \frac{n}{R}\pm \frac{1}{2}mR\right) ^{2}=\frac{n^{2}}{2R^{2}}+\frac{m^{2}R^{2}}{8}\pm \frac{nm}{2}\]
They are created by the vertex operators \( V_{n,m} \), where \( n\in \mathbf{R} \)and the winding number \( m \)
(which is the eigenvalue of the topological charge \( Q \)) satisfies \( m\in \mathbf{Z} \). \textit{Locality}
of the operator product algebra puts a severe constraint on the operator content
of the UV field theory. At a generic value of the coupling constant \( \beta  \) \textit{there
are only two different possible maximal local subalgebras} of the algebra generated
by the vertex operators \cite{kl-me}, namely a bosonic algebra \( A_{b} \) generated by operators
with \( \{n,\: m\in \mathbf{Z}\} \) which corresponds to the modular invariant partition function and is
the UV limit of the sG model, and a fermionic algebra \( A_{f} \) generated by operators
with \( \{n\in \mathbf{Z},\: m\in 2\mathbf{Z}\}\bigcup \{n\in \mathbf{Z}+\frac{1}{2},\: m\in 2\mathbf{Z}+1\} \) corresponding to the \( \Gamma _{2} \) invariant partition function and the UV limit
of mTh field theory. The two algebras overlap in the sector with even winding
number (topological charge) which is the sector common in the two off-critical
field theories and is the one that so far has been studied using the NLIE. We
limit ourselves to this sector in this paper too.

The ultraviolet behaviour of the solutions of the NLIE can be obtained in the
so-called ``\textit{kink limit}'' well-known from TBA. The method of calculation
is the same as outlined in \cite{noi}\cite{ddv-97} so let us just give the results. For the vacuum
(quantized with \textit{half-integer} Bethe quantum numbers) one gets the leading
behaviour

\[
P^{\pm }=-\frac{\pi c}{6L}+...\: ,\: L\rightarrow 0\]
with \( c=1 \), which is the correct UV central charge. 

For a two-hole state \textit{quantized with integers} (which is the prescription
of \cite{ddv-97}) and having a left-moving hole with quantum number \( I^{+}\geq 0 \) and a right-moving
one with quantum number \( I^{-}\leq 0 \) (with \( I^{+}\neq I^{-} \)) we obtain
\[
\Delta ^{\pm }=\frac{1}{8R^{2}}+\frac{1}{2}R^{2}-\frac{1}{2}\pm I^{\pm },\]
which corresponds to the vertex operators \( V_{\pm \frac{1}{2},2} \) and their descendants. As these
operators \textit{are not included in any of the two possible maximal local
operator algebras}, we have to exclude the integer-quantized two-hole states.
The UV dimensions of the same two-hole configuration with \textit{half-integer
quantization} turns out to be
\[
\Delta ^{\pm }=\frac{1}{2}R^{2}-\frac{1}{2}\pm I^{\pm },\]
which instead describes \( V_{0,2} \) and its descendants, the primary state corresponding
to the minimal choice \( I^{\pm }=\pm \frac{1}{2} \). This is in fact the vertex operator that one expects
to be the UV limit of the lowest lying two-soliton state in view of the results
in \cite{kl-me}. The results presented in \cite{ddv-97} are different when specialised to the two-hole
states. Since (as shown in the next section) the numerical results confirm our
calculations we think that the calculation of the ultraviolet dimensions in
\cite{ddv-97} should be taken with care.

For generic states with even number of solitons \( H \), half of which move to the
left and half to the right, with their quantum numbers not necessarily symmetrical
to the origin, choosing \textit{half-integer quantization} we obtained that
the UV limit corresponds to the family \( V_{0,H} \) (the primary state given by the lowest
possible choice of Bethe quantum numbers). This contradicts the rule in \cite{ddv-97} which
is integer quantization for \( H/2 \) odd and half-integer for \( H/2 \) even, but is in accordance
with what is expected from the UV conformal field theory and agrees with the
numerical results we present in the next section. Integer quantization, instead,
leads to the vertex operators \( V_{\pm \frac{1}{2},H} \) and their descendants, which are again excluded
by locality.

\section{\label{tcsa_section}Truncated Conformal Space at \protect\( c=1\protect \)}

In this section we describe results obtained using the Truncated Conformal Space
(TCS) method which support the validity of the NLIE describing the vacuum and
pure hole states.

The TCS method was originally created to describe perturbations of Virasoro
minimal models in finite spatial volume \cite{yurzam}. We have developed an extension of
the method to study perturbations of a \( c=1 \) compactified boson, more closely the
perturbation corresponding to sG theory.

The perturbation keeps the winding number \( m \) and the momentum \( P \) conserved. Therefore,
the Hilbert space of the massive theory can be split into sectors labelled by
the values of \( P \) and \( m \), which are quantized by integers. The TCS method consists
of retaining only those states in such a sector for which the eigenvalue of
the UV conformal Hamiltonian \( H_{CFT} \) is less than a certain upper value \( E_{cut} \), so the
truncated space is defined as

\[
{\cal H}_{TCS}(s,m,E_{cut})=\left\{ \left| \Psi \right\rangle :P\left| \Psi \right\rangle =s\left| \Psi \right\rangle ,Q\left| \Psi \right\rangle =m\left| \Psi \right\rangle ,H_{CFT}\left| \Psi \right\rangle \leq E_{cut}\left| \Psi \right\rangle \right\} \: .\]

For a given value of \( s \), \( m \) and \( E_{cut} \) this space is always finite dimensional. In
this space, the Hamiltonian is represented by a finite size matrix whose explicit
form is:

\[
\widehat{H}=\frac{2\pi }{L}\left( \widehat{L_{0}}+\widehat{\overline{L}_{0}}-\frac{c}{12}\widehat{I}+g\frac{L^{2-2\Delta _{V}}}{\left( 2\pi \right) ^{1-2\Delta _{V}}}\widehat{B}\right) \: ,\]
where \( \widehat{L_{0}} \) and \( \widehat{\overline{L_{0}}} \) are diagonal matrices with their diagonal elements being the left
and right conformal weights, \( \widehat{I} \) is the identity matrix and the matrix elements
of \( \widehat{B} \) are

\[
\widehat{B}_{\Phi ,\Psi }=\left\langle \Phi \right| :\cos (\hat{\beta }\varphi (1,1)):\left| \Psi \right\rangle \: .\]
We choose our energy units in terms of the soliton mass \( M \), related to the coupling
constant \( g \) by the mass gap formula given in ref. \cite{mass_scale}:

\[
g=\kappa M^{2-2\Delta _{V}}\quad ,\quad \kappa =\frac{2\Gamma (\Delta _{V})}{\pi \Gamma (1-\Delta _{V})}\left( \frac{\sqrt{\pi }}{2\Gamma \left( \frac{1}{2-2\Delta _{V}}\right) \Gamma \left( \frac{h}{4-4\Delta _{V}}\right) }\right) ^{2-2\Delta _{V}}.\]

The TCS method provides a nonperturbative way to obtain the spectrum (the mass
gap, the mass ratios and the scattering amplitudes) numerically which can serve
as a tool to check the predicted exact results of integrable field theories
and get a picture of the physical behaviour even for the non-integrable case.
The systematic error introduced by the truncation procedure is called the \textit{truncation
error}. It increases with the volume \( L \) and can be made smaller by increasing
the truncation level (at the price of increasing the size of the matrices, which
is bound from above by machine memory and computation time).

While we do not describe technical details of the TCS computation (we refer
the interested reader to \cite{yurzam}), we do have to make some remarks on how the TCS
method applies to \( c=1 \) theories. 

The (dimensionless) energy of any state \( \Psi  \) (the \textit{scaling function}) goes
with the volume \( L \) as

\begin{equation}
\label{scalingfun}
\frac{E_{\Psi }(L)}{M}=-\frac{\pi \left( c-12\left( \Delta ^{+}_{\Psi }+\Delta ^{-}_{\Psi }\right) \right) }{6l}+Bl+\sum ^{\infty }_{k=1}C_{k}\left( \Psi \right) l^{k(2-2\Delta _{V})}\: ,
\end{equation}
where \( \Delta _{\Psi }^{\pm } \) are the left/right conformal dimensions of the state in the ultraviolet
limit, \( B \) is the universal bulk energy constant (the vacuum energy density) and
the infinite sum represents the perturbative contributions from the potential
\( V \). 

The bulk energy constant has been predicted from TBA and reads \cite{mass_scale}

\begin{equation}
\label{bulk}
B=-\frac{1}{4}\tan \left( \frac{p\pi }{2}\right) 
\end{equation}
(the same was obtained from the NLIE in \cite{ddv-95}). This is a highly non-analytic function
of \( p \) and it becomes infinite at the points where \( p \) is an odd integer. In fact,
at these points there is a value of \( k \) for which \( k(2-2\Delta _{V})=2 \), and \( C_{k}\left( \Psi \right) \: \rightarrow \: \infty  \). The infinite parts
of \( B \) and \( C_{k}\left( \Psi \right)  \) exactly cancel, leaving a logarithmic (proportional to \( l\: \log l \)) and a finite
linear contribution to the energy, by a sort of a resonance mechanism. 

In the attractive regime the results of TCS converge very well when increasing
the truncation level \( E_{cut} \) (as far as we do not approach too much the free fermion
point \( p=1 \)). On the contrary, in the repulsive regime there are convergence problems
for the scaling functions \( E_{\Psi }(L) \). These are related to certain ultraviolet divergences
in conformal perturbation theory already known in the literature (for a detailed
discussion see \cite{kl-me2}), which disappear when we subtract the vacuum energy from the
excited states energies (which is a sort of a renormalization procedure). The
\textit{relative} energies \( {\cal E}_{\Psi }(L)=E_{\Psi }(L)-E_{vac}(L) \) converge well, which can be seen e.g. from the
fact that the mass gap acquires its exact value (which is \( 1 \) in our units) quite
rapidly when increasing the truncation level. Consequently, in the repulsive
regime one can only trust the \textit{relative} scaling functions produced by
the TCS method, while in the attractive regime we will see that even the \textit{absolute}
energy values agree very well with the predictions of the NLIE (including the
predicted bulk energy constant (\ref{bulk}) , which is completely analytic for \( p<1 \) and
thus logarithmic corrections are absent as well). 

Even in the attractive regime, to get reliable results we had to work with dimensions
around \( 4000 \), which was just around the limitation of our machines. Diagonalization
for one sector took around one day for twenty evenly spaced values of the volume
\( L \). This shows that the TCS for \( c=1 \) theories is far less convergent than the one
for minimal models (in the original Yang-Lee example the authors of \cite{yurzam} took a
\( 17 \) dimensional Hilbert space (!) and arrived to very accurate results). Such
a slow convergence meant that the total procedure (building up the basis, calculating
the matrix elements and diagonalizing the Hamiltonian) had to be done completely
in a compiled C program. For the matrix diagonalization, we used the LAPACK
numerical linear algebra library, freely available from Netlib over the Internet.

\section{Numerical results}

\subsection{TCS and NLIE in the attractive regime}

We have made the comparison of the numerical data from TCS and the NLIE predictions
at several values of the parameter \( p \). For illustration, we present the case
of \( p=\frac{2}{7} \). For this and all other values of the coupling constant, we obtained a
spectacular agreement between the results obtained by the two methods, up to
deviations of order \( 10^{-4}-10^{-3} \) which are smaller than the size of the points in the plots.
The deviation grows with the volume \( L \), exactly as expected for truncation errors.
By studying other parameters such as the mass gap, the breather-soliton mass
ratios and the rate of convergence of the energy levels with increasing the
value of \( E_{cut} \), the small differences can be clearly attributed to the inaccuracy
in the TCS data. 

Figures 1, 2 and 3 show the results in the \( m=0,\: 2 \) and \( 4 \) sectors, respectively. The
value of the predicted bulk energy (\ref{bulk}) has been subtracted from TCS data in
order to normalize them the same way as the NLIE data are. Note that in order
to achieve agreement the holes should be quantized with half-integers. We have
made some trials of integer quantization in the NLIE, but the results obtained
differed significantly from the TCS data, with errors of the order \( 10\: \% \). 

\begin{figure}
{\centering \resizebox*{0.7\textwidth}{0.35\textheight}{\includegraphics{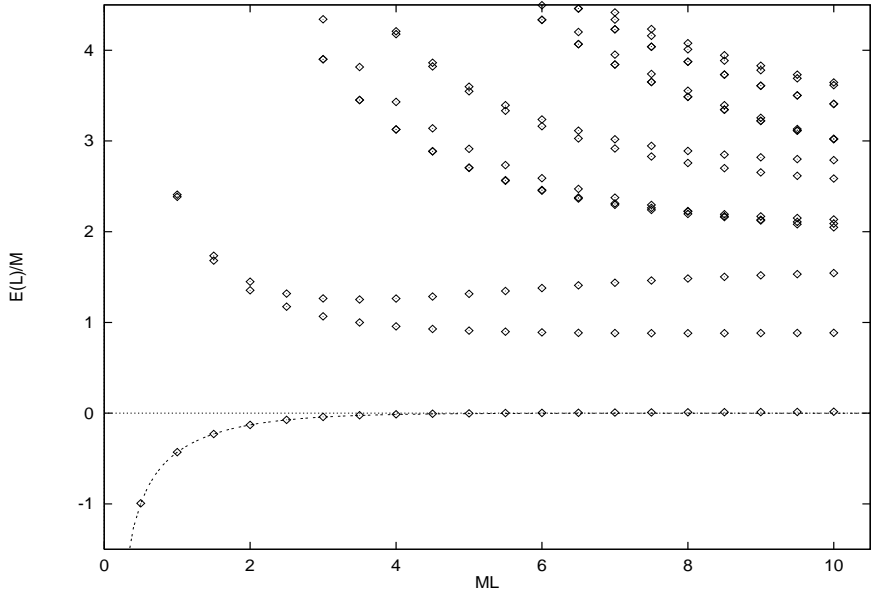}} \par}

\caption{The first few energy levels in the vacuum (\protect\( m=0\protect \)) sector at \protect\( p=\frac{2}{7}\protect \) (plotted with diamonds)
for \protect\( E_{cut}=17.0\protect \) (the dimension of the space is \protect\( 4141\protect \)) and the NLIE prediction for the vacuum
scaling function (shown with a continuous line).}
\end{figure}

\begin{figure}
{\centering \resizebox*{0.7\textwidth}{0.35\textheight}{\includegraphics{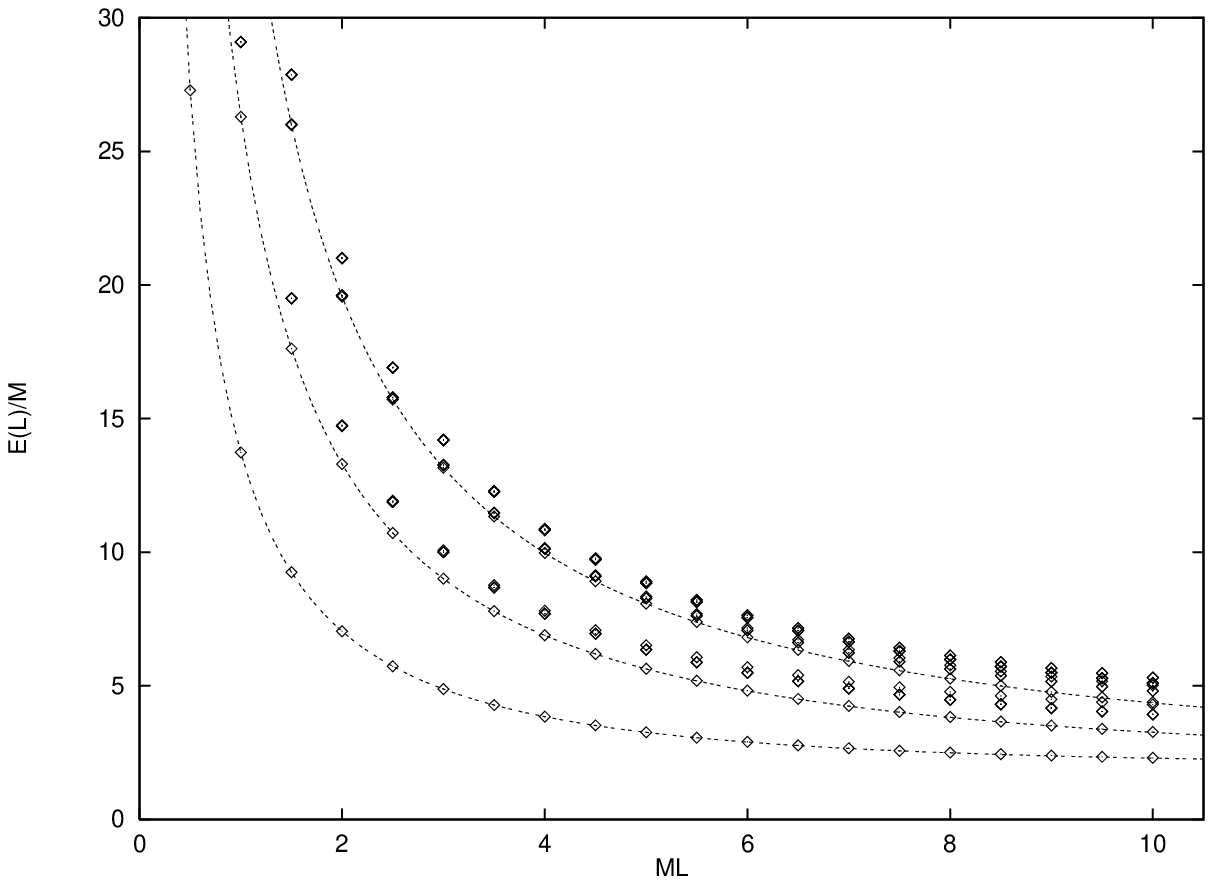}} \par}

\caption{The first few energy levels in the \protect\( m=2\protect \) sector at \protect\( p=\frac{2}{7}\protect \) (plotted with diamonds) for
\protect\( E_{cut}=20.0\protect \) (the dimension of the space is \protect\( 3917\protect \)) and the NLIE prediction for the two hole
scaling functions with quantum numbers \protect\( \left( -\frac{1}{2},\frac{1}{2}\right) \protect \), \protect\( \left( -\frac{3}{2},\frac{3}{2}\right) \protect \) and \protect\( \left( -\frac{5}{2},\frac{5}{2}\right) \protect \) (shown with continuous lines).}
\end{figure}\begin{figure}
{\centering \resizebox*{0.7\textwidth}{0.35\textheight}{\includegraphics{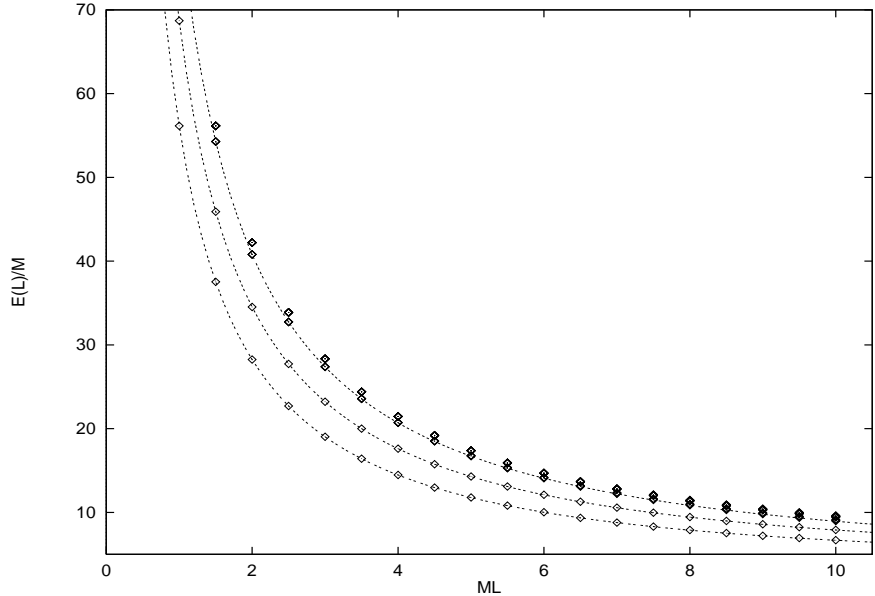}} \par}

\caption{The first few energy levels in the \protect\( m=4\protect \) sector at \protect\( p=\frac{2}{7}\protect \) (plotted with diamonds) for
\protect\( E_{cut}=26.0\protect \) (the dimension of the space is \protect\( 3403\protect \)) and the NLIE prediction for the four hole
scaling functions with quantum numbers \protect\( \left( -\frac{3}{2},-\frac{1}{2},\frac{1}{2},\frac{3}{2}\right) \protect \), \protect\( \left( -\frac{5}{2},-\frac{1}{2},\frac{1}{2},\frac{5}{2}\right) \protect \) and \protect\( \left( -\frac{5}{2},-\frac{3}{2},\frac{3}{2},\frac{5}{2}\right) \protect \) (shown with continuous lines).}
\end{figure}

\subsection{The repulsive regime}

In the repulsive regime the \textit{absolute} scaling functions do not converge
well in the TCS method as it was described in Section 4 above. We take as an
example the value of the coupling \( p=1.5 \) and plot the \textit{relative} scaling functions
obtained from TCS, comparing them with the corresponding ones obtained from
the NLIE (figures 4 and 5). One can see that the truncation errors become larger,
at values of the volume \( l \) close to \( 10 \) the deviation can be observed even from
the figures. For \( l<5 \) the agreement is still within an error of order \( 10^{-3} \). Once again,
results obtained with integer quantization for the holes showed a large deviation
from the TCS data. 

\begin{figure}
{\centering \resizebox*{0.7\textwidth}{0.35\textheight}{\includegraphics{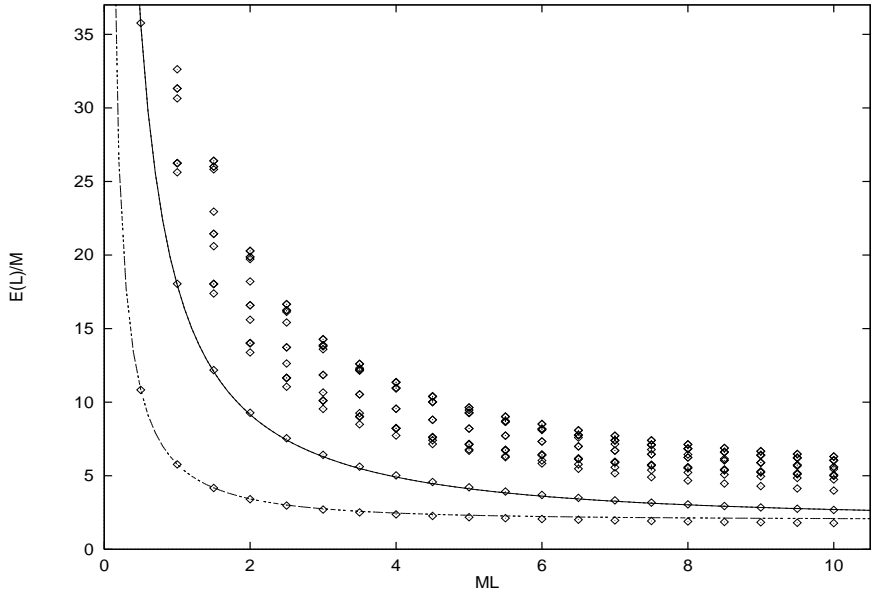}} \par}

\caption{The first few energy levels relative to the vacuum in the \protect\( m=2\protect \) sector at \protect\( p=1.5\protect \) (plotted
with diamonds) for \protect\( E_{cut}=20.0\protect \) (the dimension of the space is \protect\( 4445\protect \)) and the NLIE prediction
for the relative two hole scaling functions with quantum numbers \protect\( \left( -\frac{1}{2},\frac{1}{2}\right) \protect \) and \protect\( \left( -\frac{3}{2},\frac{3}{2}\right) \protect \) (shown
with continuous lines). }
\end{figure}

\begin{figure}
{\centering \resizebox*{0.7\textwidth}{0.35\textheight}{\includegraphics{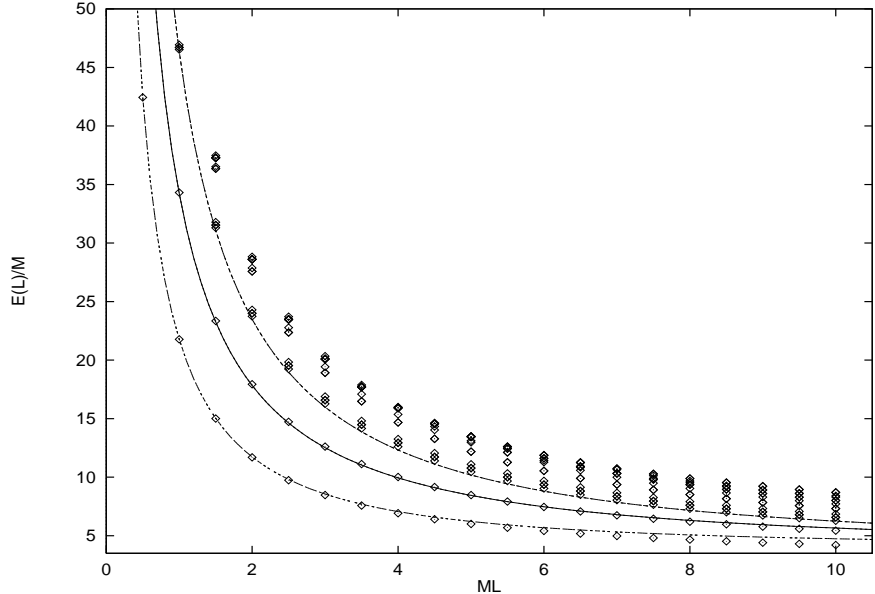}} \par}

\caption{The first few energy levels relative to the vacuum in the \protect\( m=4\protect \) sector at \protect\( p=1.5\protect \) (plotted
with diamonds) for \protect\( E_{cut}=22.5\protect \) (the dimension of the space is \protect\( 4149\protect \)) and the NLIE prediction
for the relative two hole scaling functions with quantum numbers \protect\( \left( -\frac{3}{2},-\frac{1}{2},\frac{1}{2},\frac{3}{2}\right) \protect \), \protect\( \left( -\frac{5}{2},-\frac{1}{2},\frac{1}{2},\frac{5}{2}\right) \protect \) and \protect\( \left( -\frac{5}{2},-\frac{3}{2},\frac{3}{2},\frac{5}{2}\right) \protect \)
(shown with continuous lines). }
\end{figure}

\section{Conclusions}

This letter should be seen as a preliminary report of the work done by our group
to better investigate and clarify some of the properties of the NLIE in sG theory.
We have given the first numerical check against Truncated Conformal Space data
-- the only checks so far have been calculations of the conformal spectrum at
small scale (kink limit) and comparison with the Factorized Scattering Theory
at large scale. The results presented here are the first evidence of the validity
of NLIE in the \textit{intermediate range of scales} and thus show that the
NLIE describes correctly the interpolation between the UV and IR asymptotic
regions. Our approach has also resolved some ambiguities about the choice of
quantization rule for the Bethe quantum numbers: \emph{pure hole states are
always quantized by half-integers} in the sG model on a cylinder with periodic
boundary conditions. Work is in progress to extend the results to complex roots,
the outcome of which will be published -- together with details omitted in this
paper -- in a forthcoming article \cite{noi3}.

\medskip{}
\textbf{Acknowledgements --} We are indebted to C. Destri, V. A. Fateev and
E. Quattrini for useful discussions. This work was supported in part by NATO
Grant CRG 950751, by European Union TMR Network FMRX-CT96-0012 and by INFN \emph{Iniziativa
Specifica} TO12. G. T. has been partially supported by the FKFP 0125/1997 and
OTKA T016251 Hungarian funds.

\end{document}